\documentclass[a4paper]{jpconf}
\usepackage{amsmath}
\usepackage{amssymb}
\usepackage{graphicx,wrapfig,psfig}
\usepackage{color}

\begin{document}
\renewcommand{\vec}[1]{\boldsymbol{#1}}
\newcommand{\tens}[1]{\mathsf{#1}}
\newcommand{\pd}[2]{\frac{\partial #1}{\partial #2} }
\newcommand{\HALF}{\frac{1}{2}}
\newcommand{\vol}{\mathcal{V}}
\newcommand{\DS}{\displaystyle}
\newcommand{\fpath}{./Figs/}
\def\p{\mathrm{p}}
\def\e{\mathrm{e}}
\def\I{\mathrm{I}}

\newcommand{\aap}{A\&A}
\newcommand{\apj}{Astrophys. J.}
\newcommand{\apjs}{Astrophys. J. Suppl.}
\newcommand{\mnras}{MNRAS}
\newcommand{\apjl}{Astrophys. J. Lett.}
\newcommand{\ssr}{Space Sci. Rev.}
\newcommand{\apss}{Astrophys. Space Sci.}
\newcommand{\nat}{Nature}
\newcommand{\jgr}{J. Geophys. Res. Space Phys.}
\newcommand{\grl}{Geophys. Res. Lett.}
\newcommand{\solphys}{Solar Phys.}

\title{Pickup Ion Effect of the Solar Wind Interaction with the Local Interstellar Medium}

\author{N.~V.~Pogorelov,$^1$ M.~C.~Bedford,$^1$ I.~A.~Kryukov,$^{2,1}$ and G.~P.~Zank$^1$}

\address{$^1$Department of Space Science, University of Alabama in Huntsville, Huntsville, AL 35805, U.S.A.}
\address{$^2$Institute for Problems in Mechanics, Russian Academy of Sciences, Moscow, Russia}
\ead{nikolai.pogorelov@uah.edu}

\begin{abstract}
Pickup ions are created when interstellar neutral atoms resonantly exchange charge with the solar wind (SW) ions, especially
in the supersonic part of the wind, where they carry most of the plasma pressure. Here we present numerical simulation results
of the 3D heliospheric interface treating pickup ions as a separate proton fluid. To satisfy the fundamental conservation laws, we
solve the system of equations describing the flow of the mixture of electrons, thermal protons, and pickup ions.
To find the density and pressure of pickup ions behind the termination shock, we employ simple boundary conditions that take
into account the \emph{Voyager} observations that showed that the decrease in the kinetic energy of the mixture at the termination shock predominantly contributed to the increase in the pressure of pickup ions. We show that this model adequately describes the flow
of the plasma mixture and results in a noticeable decrease in the heliosheath width.
\end{abstract}

\section{Introduction}
The solar wind (SW) interaction with the local interstellar medium (LISM) is strongly affected by charge exchange
between ions, predominantly protons, and neutral atoms, predominantly neutral hydrogen (H). The importance of charge exchange had been
acknowledged before the first quantitative model of the SW--LISM interaction was put forward  \cite{1969Natur.223..936B,1971NPhS..233...23W,1975Natur.254..202W,1977Holzer}. The reason for this is based on the partial ionization
of the LISM. In particular, the neutral H density in the LISM is about three times greater than the proton density.
In principle, there is no direct measurement of latter density because the plasma instrument onboard \emph{Voyager} 1 (\emph{V}1), which is
believed to have been traversing the LISM since July 2012 \cite{Stone-etal-2013,Burlaga147,Krimigis-etal-2013,Webber-McDonald-2013,Burlaga-Ness-2014}. The electron density has been inferred to be about 0.08 cm$^{-3}$ from the plasma wave instrument observations \cite{Gurnett-etal-2013}. The neutral H density in the unperturbed LISM is usually derived
from numerical simulations, in particular, to satisfy the condition that the H density be 0.08--0.09 cm$^{-3}$ in the inner heliosphere.
These values are related to pickup ion (PUI) measurements by \textit{Ulysses} \cite{2009SSRv..143..163G,2009SSRv..143..177B} and SW ion deceleration at \emph{Voyager} 2 (\emph{V}2) due to charge exchange \cite{2008A&A...491....1R}. PUIs are secondary ions that are born when
primary ions experience charge exchange with neutral atoms. As PUIs born from different H atoms in different regions of the SW--LISM interaction have substantially different properties, they should ideally be modeled as multiple populations, e.g., as in \cite{Malama}, and kinetically.
The SW--LISM interface is usually, for convenience of interpretation, subdivided into 4 regions: (1) region~0 is in the LISM unperturbed by the presence of the heliosphere; (2) region~1 is the LISM plasma region modified due to the flow deceleration at the heliopause (HP) -- a tangential discontinuity which separates the SW from the LISM, if ideal MHD terminology is used; (3) region~2 is between the HP and the heliospheric termination shock (TS); and (4) region~3 is the supersonic SW.
A PUI created in any of these regions experiences the action of an electric field that acts on it in the plasma frame until the PUI bulk velocity
becomes equal to the plasma bulk velocity \cite{Parker-1965}. The newly created PUIs form essentially a ring-beam distribution with the speed
ranging between 0 and 2$V_\mathrm{SW}$ in the Sun's inertial frame \cite{1976Vsyliunas}.  Such distribution function is unstable to a number of instabilities
(see \cite{1976Vsyliunas} and references therein), which results in a spherical shell distribution that becomes filled as the SW propagates outwards \cite{Isenberg87,1994JGR....9919229W}. The heliosphere beyond the ionization cavity is dominated thermally by PUIs \cite{Burlaga_etal_1994,Richardson_etal_1995a,Zank-1999,2014ApJ...797...87Z,Zank-2015}.
According to~\cite{Decker_etal_2008,Decker_etal_2015}, the inner heliosheath (IHS) pressure contributed
by energetic PUIs and anomalous cosmic rays far exceeds that of the thermal background plasma and magnetic field. The IHS here coincides with region~2 introduced above.

A number of one-plasma-fluid models \cite{1993JGR....9815157B,Pauls95,2006ApJ...644.1299P,Jacob06,Jacob07,2005A&A...437L..35I,2009SSRv..146..329I,Opher-etal-2006} take into account the effect PUIs by assuming that they are in thermal equilibrium with the background plasma. Although this is not true, the conservation laws of mass, momentum, and energy for
the mixture of electrons, ions, and PUIs are still satisfied approximately. According to \cite{2014ApJ...797...87Z,Zank-2015}, such approaches may be correct  only if the heat conduction tensor and the dissipation coefficient due to ion-PUI interaction are both zero. This happens when PUIs are completely coupled with the thermal ions and the scattering time is infinitely small.
It was shown in~\cite{Isenberg_1986} that the effect of PUIs can be quantified if they are treated as a separate fluid. In \cite{2006JGRA..111.7101U,2012ApJ...754...40U,2012AIPC.1436...48K,2011JGRA..116.3105D}, time-dependent simulations of the supersonic SW are presented with PUIs treated as a separate fluid and turbulence effects taken into account.
These models still make the approximation of instantaneous isotropization, i.e., that the PUI description can be described as a filled shell, thus neglecting the nearly-isotropic character of the PUI distribution function when a finite scattering time is included \cite{2014ApJ...797...87Z}.

In principle, any PUI fluid model is an approximation of their kinetic behavior. Such a model was developed by \cite{Malama}
on the assumptions that PUIs are isotropic away from discontinuities and the TS is a perpendicular shock. However, proper boundary conditions are necessary to describe the PUI crossing the TS. Paper~\cite{Gamayunov} investigates the evolution of the PUI  distribution function (in a pitch-angle-averaged approximation) together with the PUI-generated waves PUIs that heat the thermal SW ions.

The extension of a 2-fluid SW model developed for the supersonic flows is not straightforward. This is because the pressure equation for PUIs used in such models is not valid across the TS. Instead, some boundary conditions should be developed for the PUI density and pressure, $\rho_\mathrm{PUI}$ and $p_\mathrm{PUI}$. Such boundary conditions are discussed in \cite{1996SoPh..168..389C,1996JGR...101..457Z,2008A&A...490L..35F,2010ApJ...708.1092Z,2012Ap&SS.341..265F,2013A&A...558A..41F}.
It is known, e.g., the most of the SW kinetic energy was absorbed by PUI when the TS was crossed by \textit{V}1 \cite{Richardson_etal_2008Nature} (see also predictions in \cite{1996JGR...101..457Z}). One would not expect a simple finite approximation of $p_\mathrm{PUI}$ across the TS to satisfy such boundary conditions. In principle, as explained, e.g., in \cite{Kulik1}, one should either write out and solve a system of conservation laws across of a discontinuity or specify the boundary conditions at it. These approaches are equivalent, because the number of possible conservation laws is infinite, which results in an infinite number of shock boundary conditions. It is understood that only the fundamental physical conservation laws are appropriate for the description of
magnetized fluid flows. In contrast, \cite{Usmanov16} solve non-conservative pressure equations everywhere in the SW--LISM interaction region, which violates the basic principles required to solve systems of hyperbolic equations. Moreover, the turbulence model of \cite{Breech_2008} used in \cite{Usmanov16} is based on an Alfv\'en-mode approximation and is invalid in the IHS. Moreover, the source terms used in \cite{Usmanov16} are valid only for ``cold'' plasma and are not applicable to the hot IHS plasma.

In this paper, we use simple boundary conditions at the TS which allows us to demonstrate that the IHS width decreases when the PUI fluid is treated separately.

\section{Governing Equations}
	We present the system of equations that governs the flow of the plasma mixture (thermal protons, PUIs, and electrons),
neutral atom populations, and PUIs. Subscripts for charge exchange source terms are identified by the subscripts $a$, which is a fixed integer for the corresponding population of neutral atoms, $\p$ for thermal protons, and $\I$ for pickup protons:

	\subsection*{Plasma mixture}
		\begin{eqnarray}
		&&\frac{\partial\rho}{\partial t}+\nabla\cdot(\rho\mathbf{u}) = 0 = D_\Sigma, \\
		&&\frac{\partial\rho\mathbf{u}}{\partial t}+\nabla\cdot\left(\rho\mathbf{u}\mathbf{u}+p^*\mathbf{I}-\frac{1}{4\pi}\mathbf{B}\mathbf{B}\right) = \mathbf{M}_\Sigma, \\
		&&\frac{\partial E}{\partial t}+\nabla\cdot\left((E+p^*)\mathbf{u}-\frac{1}{4\pi}(\mathbf{B}\cdot\mathbf{u})\mathbf{B}\right) = E_\Sigma, \\
		&&\frac{\partial\mathbf{B}}{\partial t}+\nabla\cdot(\mathbf{u}\mathbf{B}-\mathbf{B}\mathbf{u}) = 0.
	\end{eqnarray}
	\subsection*{Neutrals}
		\begin{eqnarray}
		&&\frac{\partial\rho_a}{\partial t}+\nabla\cdot(\rho_a\mathbf{u}_a)= D_a, \\
		&&\frac{\partial\rho_{a}\mathbf{u}_a}{\partial t}+\nabla\cdot(\rho_a\mathbf{u}_a\mathbf{u}_a+p_a\mathbf{I})= \mathbf{M}_a, \\
		&&\frac{\partial e_a}{\partial t}+\nabla\cdot((e_a+p_a)\mathbf{u}_a)= E_a.
	\end{eqnarray}
	\subsection*{Pick-up ions}
		\begin{eqnarray}
		&&\frac{\partial\rho_\I}{\partial t}+\nabla\cdot(\rho_\I\mathbf{u})= D_\I, \\
		&&\frac{\partial p_\I}{\partial t}+\nabla\cdot(p_\I\mathbf{u})+(\gamma-1)p_\I\nabla\cdot\mathbf{u}= P_\I
	\end{eqnarray}

Here $p$ is the pressure, $\rho$ the mass density, $\mathbf{u}$ the bulk velocity vector, $\mathbf{B}$ the magnetic filed vector, $e$ the total
energy density (includes the magnetic energy), $p^*$ the total pressure of the mixture (including magnetic energy), and $\mathbf{I}$ is the identity tensor. The source terms $D$, $\mathbf{M}$, $E$,a nd $P$ are defined later. The subscript $\Sigma$ refers to the mixture properties.

The following quantities are used in the definition of the source terms \cite{Zank96}:
\begin{eqnarray}
			U_{ij}^\rho&=&\left[\frac{4}{\pi}(v_{T_i}^2+v_{T_j}^2)+(\mathbf{u}_i-\mathbf{u}_j)^2\right]^{1/2}, \\
			U_{ij}^m&=&\left[\frac{16}{\pi}v_{T_i}^2+\frac{9\pi}{4}v_{T_j}^2+4(\mathbf{u}_i-\mathbf{u_j})^2\right]^{1/2}, \\
			U_{ij}^e&=&\left[\frac{4}{\pi}v_{T_i}^2+\frac{64}{9\pi}v_{T_j}^2+(\mathbf{u}_i-\mathbf{u}_j)^2\right]^{1/2}, \\
			\lambda_{ij}(U_{ij}^\rho)&=&\sigma_{\mathrm{cx}}(U_{ij}^\rho)=\left(13.493-0.531\log U^\rho_{ij}\right)^2\left(1-\exp\frac{-2.94\times10^9}{U^\rho_{ij}}\right)^{4.5}
\end{eqnarray}

Here the subscripts $i$ and $j$ can have values of $a$ and either $\p$ or $\I$. The quantity $v_T$ is the thermal velocity and $\lambda_{ij}$
is the charge exchange cross section of particles of type $i$ with particles of type $j$.

The source terms for the density, momentum, energy, and PUI pressure equations have the form
\begin{eqnarray}
		&&H_{ak}^\rho=\lambda_k\rho_k\rho_aU_{ak}^\rho, \qquad k=\p,\I \\
		&&\mathbf{H}_{ak}^m=\lambda_k\rho_k\rho_a\left(U_{ak}^\rho\mathbf{u}+\frac{v_{T_k}^2}{U_{ak}^m}(\mathbf{u}-\mathbf{u}_a)\right), \\
		&&\mathbf{H}_{ka}^m=\lambda_k\rho_k\rho_j\left(U_{ka}^\rho\mathbf{u}_a-\frac{v_{T_a}^2}{U_{ka}^m}(\mathbf{u}-\mathbf{u}_a)\right), \\
		&&H_{ak}^e=\lambda_k\rho_k\rho_a\left(\frac{1}{2}u^2U_{ak}^\rho+
\frac{3}{4}v_{T_k}^2U_{ak}^e+\frac{v_{T_k}^2}{U_{ak}^m}\mathbf{u}\cdot(\mathbf{u}-\mathbf{u}_a)\right),\\
&&H_{ka}^e=\lambda_k\rho_k\rho_a\left(\frac{1}{2}u^2_aU_{ka}^\rho+
\frac{3}{4}v_{T_a}^2U_{ka}^e-\frac{v_{T_a}^2}{U_{ka}^m}\mathbf{u}_a\cdot(\mathbf{u}-\mathbf{u}_a)\right),\\
		&&H_{ak}^p=\lambda_k\rho_k\rho_a(\gamma-1)\frac{3}{4}v_{T_k}^2U_{ak}^e, \\
&&H_{ka}^p=\lambda_k\rho_k\rho_a(\gamma-1)\left(\frac{1}{2}U_{ka}^\rho(\mathbf{u}-\mathbf{u}_a)^2+\frac{v_{T_a}^2}{U_{ka}^m}(\mathbf{u}-\mathbf{u}_a)^2+\frac{3}{4}v_{T_a}^2U_{ka}^e\right)
	\end{eqnarray}

We will assume here (this assumption can be lifted later) that there are no superthermal electrons. Thus, introducing the number density $n$,
we obtain
	\begin{equation*}
		p=n_\e kT_\e+n_\p kT_\p+n_\I kT_\I, \quad n_\e=n_\p+n_\I=n,
\end{equation*}
which reduces to
\begin{equation}
p=nkT_\e+(n-n_\I)kT_\p+n_\I kT_\I.
\end{equation}

Assuming  $T_\e=T_\p$, we obtain
\begin{equation}
\frac{p}{n}=(2-\frac{n_\I}{n})kT_\p+\frac{n_\I}{n}kT_\I.
\end{equation}

We now introduce the proton mass $m$ and thermal ion velocity $v_T^2=\frac{2kT}{m}$, which can be expressed as
\begin{equation}
v_{T_p}^2=\frac{\frac{2p}{\rho}-\frac{n_\I}{n}v_{T_\I}^2}{2-\frac{n_\I}{n}}.
\end{equation}

It is easy to notice that if $n_\I=0$, then $v_{T_p}^2=\frac{p}{\rho}$, recovering the formula for the one-ion fluid that takes into account the electron pressure. Obviously, $v_{T_a}^2=\frac{2p}{\rho}$ for neutral atoms.

	Using the results of \cite{Malama}, each ion and neutral atom is identified as a  thermal proton, pickup proton, or a neutral population atom. This is a simplification, but the best that can be done without adding more PUI fluids or using kinetic models directly.

Let us consider reactions between different species. In our extended notation, $p_0$, $p_1$, and $H_i$ stand for thermal protons, pickup protons, and neutral atoms of population $i$.  In region~3, a charge exchange results in a population 3 neutral atom and a pickup ion, except for thermal ion charge exchange with a population 3 neutral.  The latter produces a thermal proton.  So in region 3, PUIs are mostly created and thermal ions are mostly destroyed.  In region 2, we assume that PUIs are only destroyed. This is because the population of PUIs born in the IHS is substantially different from those that have crossed the TS.  Since we did not introduce additional populations of PUIs, all charge exchanges in region~2 result in thermal protons and population 2 neutrals.  For the purpose of this study, we ignore PUIs created in the outer heliosheath beyond the TS. Thus, in region 1, PUIs and thermal LISM ions are indistinguishable. All charge exchanges produce thermal protons and population 1 neutrals.
	
In the following, the subscript order and sign depend on the reaction: in region 2, for example, $\mathbf{H}^m_{\I 1}$ means the source term contribution to the momentum equation due to charge exchange of a population 1 neutral atom  with a PUI.  The formula below says that this results in a thermal ion, so it is subtracted from $\mathbf{M}_1$ and added to $\mathbf{M}_\Sigma$.  The corresponding $H^m_{1\I}$ stands for the momentum lost by a PUI when it exchanges charge with a population 1 neutral atom, so it is subtracted from $\mathbf{M}_\Sigma$ and added to $\mathbf{M}_2$.
	
The sum of all mixture and neutral source terms should be zero, but not for the PUI pressure and density.  So for region 3, the PUI source term includes positive contributions from thermal protons exchanging charge with populations 1 and 2 (which create PUIs), nothing from thermal protons exchanging charge with population 3 (no PUIs created or destroyed), both positive and negative contributions from PUIs exchanging charge with populations 1 and 2 (one PUI destroyed and another created), and only negative contributions from PUIs exchanging charge with population 3 (PUI destroyed).
	
In summary,

	\subsubsection*{Region 3}

	\begin{eqnarray*}
		&&p_0+H_1\rightarrow p_1+H_3, \quad p_1+H_1\rightarrow p_1+H_3,\\
		&&p_0+H_2\rightarrow p_1+H_3, \quad p_1+H_2\rightarrow p_1+H_3,\\
		&&p_0+H_3\rightarrow p_0+H_3, \quad p_1+H_3\rightarrow p_0+H_3.
	\end{eqnarray*}

This implies that
\[
		D_\Sigma=0,\quad D_1=-H^\rho_{\p 1}-H^\rho_{\I 1},\quad D_2=-H^\rho_{\p 2}-H^\rho_{\I 2},
\]
\[
		D_3=H^\rho_{\p 1}+H^\rho_{\p 2}+H^\rho_{\I 1}+H^\rho_{\I 2},\quad D_\I=H^\rho_{\p 1}+H^\rho_{\p 2}-H^\rho_{\I 3},
\]
\[
\mathbf{M}_\Sigma=-H^m_{1\p}+H^m_{\p 1}-H^m_{2\p}+H^m_{\p 2}-H^m_{3\p}+H^m_{\p 3}-H^m_{1\I}+H^m_{\I 1}-H^m_{2I}+H^m_{I2}-H^m_{3\I}+H^m_{\I 3},
\]
\[
		\mathbf{M}_1=-H^m_{\p 1}-H^m_{\I 1},\quad 	\mathbf{M}_2=-H^m_{\p 2}-H^m_{\I 2},
\]
\[
		\mathbf{M}_3=-H^m_{\p 3}+H^m_{3\p}+H^m_{1\p}+H^m_{2\p}-H^m_{\I 3}+H^m_{3\I}+H^m_{1\I}+H^m_{2\I},
\]
\[
	E_\Sigma=-H^e_{1\p}+H^e_{\p 1}-H^e_{2\p}+H^e_{\p 2}-H^e_{3\p}+H^e_{\p 3}-H^e_{1\I}+H^e_{\I 1}-H^e_{2\I}+H^e_{\I 2}-H^e_{3\I}+H^e_{\I 3},
\]
\[
		E_1=-H^e_{\p 1}-H^e_{\I 1},\quad E_2=-H^e_{\p 2}-H^e_{\I 2},
\]
\[
		E_3=-H^e_{\p 3}+H^e_{3\p}+H^e_{1\p}+H^e_{2\p}-H^e_{\I 3}+H^e_{3\I}+H^e_{1\I}+H^e_{2\I},
\]
\[
	P_\I=H^\p_{\p 1}+H^\p_{\p 2}+H^\p_{\I 1}-H^\p_{1\I}+H^\p_{\I 2}-H^\p_{2\I}-H^\p_{3\I}.
\]

\subsubsection*{Region 2}

	\begin{eqnarray*}
		&&p_0+H_1\rightarrow p_0+H_2, \quad p_1+H_1\rightarrow p_0+H_2,\\
		&&p_0+H_2\rightarrow p_0+H_2, \quad p_1+H_2\rightarrow p_0+H_2,\\
		&&p_0+H_3\rightarrow p_0+H_2, \quad p_1+H_3\rightarrow p_0+H_2.
	\end{eqnarray*}

This implies that
\[
	D_\Sigma=0, D_1=-H^\rho_{\p 1}-H^\rho_{\I 1},\quad D_2=H^\rho_{\p 1}+H^\rho_{\p 3}+H^\rho_{\I 1}+H^\rho_{\I 3},
\]
\[
		D_3=-H^\rho_{\p 3}-H^\rho_{\I 3},\quad 	D_\I=-H^\rho_{\I 1}-H^\rho_{\I 2}-H^\rho_{\I 3},
\]
\[
\mathbf{M}_\Sigma=-H^m_{1\p}+H^m_{\p 1}-H^m_{2\p}+H^m_{\p 2}-H^m_{3\p}+H^m_{\p 3}-H^m_{1\I}+H^m_{\I 1}-H^m_{2\I}+H^m_{\I 2}-H^m_{3\I}+H^m_{\I 3},
\]
\[
\mathbf{M}_1=-H^m_{\p 1}-H^m_{\I 1},\quad \mathbf{M}_2=-H^m_{\p 2}+H^m_{2\p}+H^m_{1\p}+H^m_{3\p}-H^m_{\I 2}+H^m_{2\I}+H^m_{1\I}+H^m_{3\I},
\]
\[
\mathbf{M}_3=-H^m_{\p 3}-H^m_{\I 3},
\]
\[
	E_\Sigma=-H^e_{1\p}+H^e_{\p 1}-H^e_{2\p}+H^e_{\p 2}-H^e_{3\p}+H^e_{\p 3}-H^e_{1\I}+H^e_{\I 1}-H^e_{2\I}+H^e_{\I 2}-H^e_{3\I}+H^e_{\I 3},
\]
\[
E_1=-H^e_{\p 1}-H^e_{\I 1},
\]
\[
	E_2=-H^e_{\p 2}+H^e_{2\p}+H^e_{1\p}+H^e_{3\p}-H^e_{\I 2}+H^e_{2\I}+H^e_{1\I}+H^e_{3\I},
\]
\[
	E_3=-H^e_{\p 3}-H^e_{\I 3},\quad P_\I=-H^\p_{1\I}-H^\p_{2\I}-H^\p_{3\I}.
\]
	
\subsubsection*{Region 1}

\[
		p_0+H_1\rightarrow p_0+H_1,\quad p_0+H_2\rightarrow p_0+H_1,\quad p_0+H_3\rightarrow p_0+H_1.
\]

This implies that
\[
	 D_\Sigma=0,\quad 	D_1=H^\rho_{\p 2}+H^\rho_{\p 3},
\]
\[
		D_2=-H^\rho_{\p 2},\quad 	D_3=-H^\rho_{\p 3},\quad 	D_\I=0,
\]
\[
		\mathbf{M}_\Sigma=-H^m_{1\p}+H^m_{\p 1}-H^m_{2\p}+H^m_{\p 2}-H^m_{3\p}+H^m_{\p 3},
\]		
\[
\mathbf{M}_1=-H^m_{\p 1}+H^m_{1\p}+H^m_{2\p}+H^m_{3\p},\quad \mathbf{M}_2=-H^m_{\p 2},\quad \mathbf{M}_3=-H^m_{\p 3},
\]
\[
		E_\Sigma=-H^e_{1\p}+H^e_{\p 1}-H^e_{2\p}+H^e_{\p 2}-H^e_{3\p}+H^e_{\p 3},
\]
\[
E_1=-H^e_{\p 1}+H^e_{1\p}+H^e_{2\p}+H^e_{3\p},\quad E_2=-H^e_{\p 2},\quad 	E_3=-H^e_{\p 3},\quad P_\I=0.
\]

The following source terms are used in the pressure equation:
\[
		H^\p_{jk}=(\gamma-1)\left(H^e_{jk}-\mathbf{u}\cdot\mathbf{H}^m_{jk}+\frac{1}{2}u^2H^\rho_{jk}\right),
\]
\[
H^\p_{ak}=(\gamma-1)\lambda_{ak}\frac{3}{4}v^2_{T_k}U^e_{ak},
\]

\section{Numerical results}
Here we present the results of our steady-state calculation only of the SW--LISM interaction. The simulation is performed
on a spherical grid with adaptive mesh refinement near the TS. The following boundary conditions have been used.
The SW is assumed to be spherically symmetric at 1~AU: the plasma density is 7.4 cm$^{-3}$, radial velocity component 450 km/s,
temperature 70,000~K, and the radial component of the magnetic field vector 49.8 $\mu$G. The heliospheric magnetic field is assumed to be
the Parker field at the inner boundary.

In the unperturbed LISM: the neutral H density is 0.172~cm$^{-3}$, plasma number density 0.082~cm$^{-3}$, velocity $V_\infty=26.4$~km/s, temperature 8,000~K, magnetic field
strength~$B_\infty=3\ \mu$G. Simulations are performed in the heliospheric coordinate system $x$, $y$, and $z$, where the $z$-axis is aligned
with the Sun's rotation axis, the $x$-axis belongs to the plane formed by the $z$-axis and the LISM velocity vector, $\mathbf{V}_\infty$, and directed upstream into the LISM, and the $y$-axis completes the right coordinate system. The directions of $\mathbf{V}_\infty$ and $\mathbf{B}_\infty$ are given by the formulae:
\[
\mathbf{V}_\infty=V_\infty (-0.996, 0, -0.089), \quad \mathbf{B}_\infty=B_\infty (0.692, -0.477, 0.541).
\]

We assumed a multi-fluid model for H atoms, as described in the Introduction. However, the extension of this approach to kinetic neutrals is
straightforward. Although the governing system of equations is written in Cartesian coordinates (and we solve it this way), the grid is spherical. This approach allows us to avoid any geometrical source terms. This is important to ensure conservation
laws hold in a flow that is characterized by the presence of
\begin{wrapfigure}{r}{85mm}
\centering
\psfig{figure=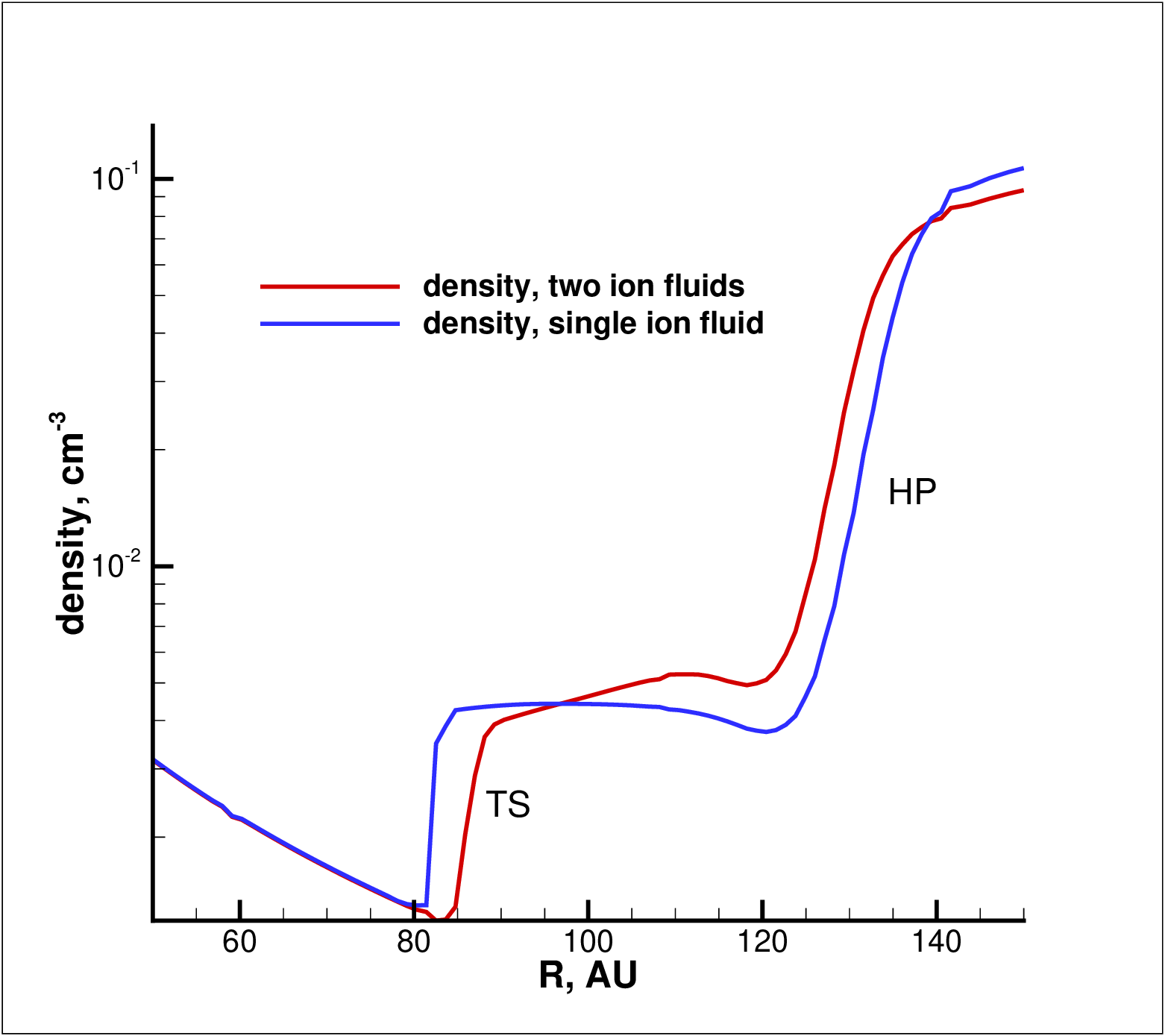,width=85mm}
\caption{\scriptsize
Ion density distributions along the \textit{V1} trajectory shows that the TS moves farther from the Sun while the HP becomes closer to it, if PUIs are treated as a separate plasma component.
}
\label{fig1}
\end{wrapfigure}
shocks and the heliopause, which our ideal-MHD approach treats as a tangential discontinuity. In principle, the presence of source terms means that the conservation laws hold only approximately, which affects the strengths of numerically-obtained discontinuities and therefore their speed.
The system of equations is solved using a Godunov-type method, where the fluxes of mass, momentum, energy, and magnetic field are obtained as solutions of an MHD Riemann problem in the linear approximation \cite{Kulik1}. The scheme has second order of accuracy in space and time.
Both MHD and Euler equations are solved in the framework of the Multi-Scale Fluid-Kinetic Simulation Suite (MS-FLUKSS) \cite{2013ASPC..474..165P,XSEDE2014}.
MS-FLUKSS is built on the Chombo adaptive mesh refinement (AMR) framework \cite{2007JPhCS..78a2013C}. The spherical volume extension of this framework was implemented for MS-FLUKSS in \cite{2008ASPC..385..197B}.
\begin{figure}
\centering
\includegraphics[width=0.48\textwidth]{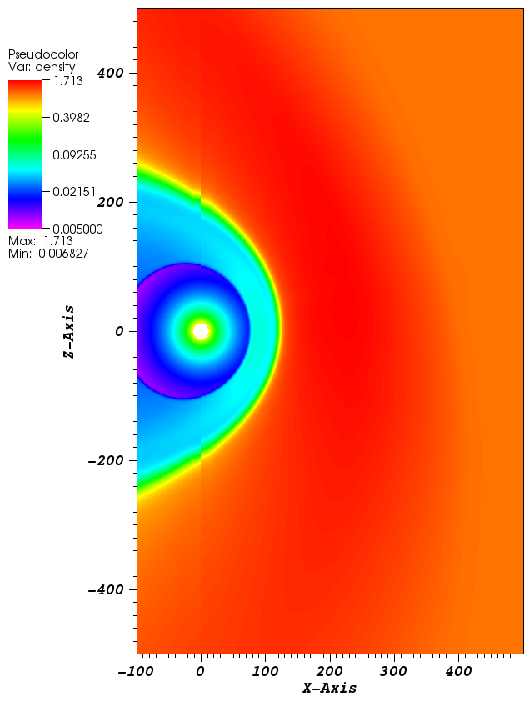}
\includegraphics[width=0.48\textwidth]{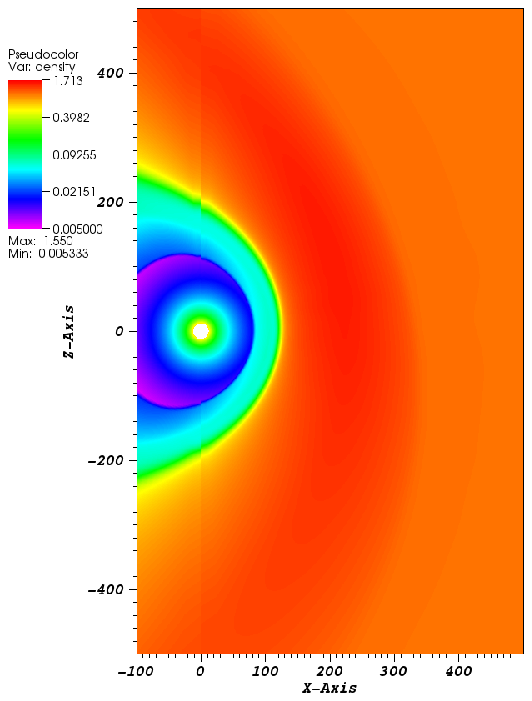}
\caption{Plasma density distributions in the meridional plane for the single-ion-fluid model (left panel) and for the case when the PUI and thermal ion fluids behave as co-moving, but distinguishable components (right panel)}.
\end{figure}

While a number of different approaches have been developed to describe the boundary conditions for PUIs across the TS which describe the energy transfer in this region \cite{2008A&A...490L..35F,2010ApJ...708.1092Z,2012Ap&SS.341..265F,2013A&A...558A..41F}, here we use a simplified approach based on the assumption that thermal protons experience only adiabatic compression, i.e., the entropy of the ion fluid is preserved.
We also assume that the density ratio behind and in front of the TS is the same for PUIs and thermal ions, so it can be determined from the corresponding ratio in the mixture. These two assumptions ensure a preferential energy influx into the PUI fluid. More accurate
approaches will be implemented in the future. Any analytic boundary conditions can be implemented in a manner similar to that presented here. To be sure that we choose proper shock conditions at the TS, or in other words, to take into account TS smearing over 2-3 computational cells, we impose these boundary conditions on the cell centers located two cells upstream and downstream.

Figure~1 shows the plasma density distribution in the direction of the Voyager~1 trajectory for the single-ion-fluid model (\emph{Case} 1) and for the case when the PUI and thermal ion fluids behave as co-moving, but distinguishable components (\emph{Case} 2). Figure~2 shows the plasma density distributions (dimensionless units with the scale of LISM plasma density) in the meridional plane (the $xz$-plane) for Case~1 (left panel) and Case~2 (right panel). It is clear from Fig.~1 that the HP moves inward towards the Sun and the TS moves outward away from the Sun with the inclusion of PUIs as a distinct component, i.e., the heliosheath width becomes quite significantly narrower when PUIs are treated distinctly from the thermal SW plasma. This result is consistent with that in \cite{Malama}
and addresses the observation made by \textit{V}1 that the HP location was closer than expected \cite{Stone-etal-2013,Burlaga147,Krimigis-etal-2013,Webber-McDonald-2013,Burlaga-Ness-2014}, one interpretation of which is that the IHS is narrower than predicted by conventional models.

\section{Conclusions}
We have performed a SW--LISM simulations based on treating of PUIs as an individual plasma component and compared our results with our ``standard'' model where PUIs are immediately assimilated with background  thermal protons. The neutral H flow was modeled using a multi-fluid approach. To make our simulations consistent with \emph{V}2 observations at the TS, we employed simple boundary conditions for the PUI pressure and density that ensured the preferential heating of PUIs across the TS. Our results show that treating PUIs as a separate plasma component indeed makes the inner heliosheath about 10~AU thinner that in a single-ion-fluid model. It should be understood, however, that while the HP was at 120~AU when V1 crossed it in August 2012, we have no information about the TS location at the same time.
It is therefore possible that the heliosheath width is not equal to 28~AU, which one would predict on assuming that neither the HP nor the TS changed their position between 2004, when V1 crossed the TS, and 2012, when it left the heliosphere and entered the LISM.
Most time-dependent SW--LISM models predict substantial variations in the TS heliocentric distance as a function of
time \cite{2000Ap&SS.274..115P,2003JGRA..108.1240Z,2005A&A...429.1069I,Pogo09,2011MNRAS.416.1475W,Pogorelov-etal-2013}.

Future work will focus on using better boundary conditions for PUIs at the TS, the implementation of a model with a kinetic treatment of H atoms, and including an extended treatment of the PUIs that goes beyond the assumption of complete isotropization.

\section*{Acknowledgments}
This work was supported, in part, by NASA grants NNX14AJ53G, NNX14AF41G, NNX14AF43G, NNX15AN72G, and NNX16AG83G, and DOE Grant DE-SC0008334. This work was also partially supported by the IBEX mission as a part of NASA's Explorer program. We acknowledge NSF PRAC award ACI-1144120 and related computer resources from the Blue Waters sustained-petascale computing project. Supercomputer time allocations were also provided on SGI Pleiades by NASA High-End Computing Program award SMD-15-5860 and on Stampede by NSF XSEDE project MCA07S033.

\section*{Bibliography}
\providecommand{\newblock}{}

\end{document}